# Social Capital and Individual Performance: A Study of Academic Collaboration

**Abstract**

Studies on social networks highlight the importance of network structure or structural properties of a given network and its impact on performance outcome. The empirical validation of the association between network structures and performance has been well documented in a number of recent studies. One of the important properties of this network structure is referred as "social capital" which is the "network of contacts" and the associated values attached to these networks of contacts. There are very few systematic empirical studies suggesting a role of co-authors, as social capital in their scientific collaboration network and their effect on performance. In this study, our aim is to provide empirical evidence of the influence of social capital and performance within the context of academic collaboration. We suggest that the collaborative process involves social capital embedded within relationships and network structures among direct co-authors. Thus, we examine whether scholars' social capital is associated with their citation-based performance, using co-authorship and citation data. In order to test and validate our proposed hypotheses, we extract publication records from Scopus having "information science" in their title or keywords or abstracts during 2001 and 2010. To overcome the limitations of traditional social network metrics for measuring the influence of scholars' social capital within their co-authorship network, we extend the traditional social network metrics by proposing a new measure (*Power-Diversity Index*). We then use Spearman's correlation rank test to examine the association between scholars' social capital measures and their citation-based performance. Results suggest that research performance of authors is positively correlated with their social capital measures. This study highlights the importance of scholars' social capital characteristics on their performance suggesting stronger links to more powerful contacts will lead to better performance and, therefore, their respective professional social network shows indicative outcomes to evaluate and predict the performance of scholars. It further highlights that the *Power-diversity Index*, which is introduced as a new hybrid centrality measure, serves as an indicator of power and influence of an individual's ability to control communication and information.

## 1. Introduction

Prominent sociologists such as Burt (1992), Coleman (1988) and Granovetter (1973) claim that personal attributes are not only the effective factor leading to the success of actors' performance, but the extent of social capital accumulated in their respective networks is more significant (Oh, Choi, & Kim, 2006). Social capital produces benefits or outcomes for individual and collective actors, which is generated through structural sources (Burt, 1992). The core idea of social capital is that a person's (or a group of people's) associates (e.g., family members, friends, colleagues) form an important asset that can be used to gain optimal performance (Woolcock & Narayan, 2000).

The concept of social capital provides a useful and comprehensive conceptual perspective (Sawyer, Crowston, & Wigand, 1999; Tsai & Ghoshal, 1998) for understanding social capital and value creation within a networking context. Accordingly, social capital means "the set of social resources embedded in relationships" (Tsai & Ghoshal, 1998, p. 464). Social capital has three components: structural, relational, and cognitive (Tsai & Ghoshal, 1998; Wellman, 1988). The structural dimension involves social interaction that the actor uses to gain access, information, or resources. The relational dimension encompasses aspects that arise from the interactions (including trust and loyalty). The cognitive dimension includes attributes such as shared norms, codes of action, and convergence of views. Our research suggests that conceptualizing social capital in terms of network structures, such as articulated by the strength of weak ties theory (Granovetter, 1973, 1983), provides valuable insight into scholars' co-authorship activities.

In most large organizations, performance of individuals and teams are measured through a set of metrics that pertain to task and contextual performance. Similarly in academia, scholars and scientists are evaluated based on their academic performance (e.g., research productivity, teaching evaluations, governance capabilities, achieved grants). Such evaluation of scholars is not only needed for faculty recruitment and promotion schemes, but also for governmental funding allocation and for achieving a high reputation within the research community. The reputation of research organizations indirectly affects society's welfare, since high reputation attracts foreign purchases, foreign investments, and highly qualified students from around the world (Abbasi, Altmann, & Hwang, 2010).

The implication of such evaluation and ranking provides the basis for governmental funding thus encouraging high research standards and goals. Therefore, on a global level, with respect to governmental funding (i.e., the allocation of funding for a specific project to a scientific research group) and university strategy, it is important to identify key scholars, collaboration areas and research strengths within universities with the aim of maximizing research output, cost optimization, and resource utilization.

Since individuals have limited capacity to acquire and use knowledge, their interactions with others are necessary for knowledge creation (Demsetz, 1991), which usually appears in the form of publishing papers. Therefore, many scientific outputs are the result of collaborative work and most research projects are too large for an individual researcher to perform alone. This, in turn, leads to large scale scientific collaboration. However, having scholars with different skills, expertise and knowledge, as human capital, in group work is needed (McFadyen, Semadeni, & Cannella, 2009). Diversity of actors involved in group work facilitates the integration of expertise, contributes to successful projects' implementation and accelerates cycle time for new product development (Cummings, 2004; Eisenhardt & Tabrizi, 1995; Griffin & Hauser, 1992; Pinto, Pinto, & Prescott, 1993), but having a basic shared understanding of each other's knowledge and expertise is necessary to have a shared understanding about the whole project or research.

The co-authorship network is a form of collaboration network among scholars that represents their scientific interactions and collective action to conduct research and produce the results as a form of a publication. Therefore, social norms and trust build among scholars, through their collaborations over time, are a form of social capital for academia. In other words, when researchers collaborate on projects, they do share substantial amounts of knowledge. This flow of knowledge, during research collaboration, becomes a stock of knowledge, which mutually benefits the researchers not only in their respective future projects (Dierickx & Cool, 1989), but also in the current research by gaining new knowledge and reputation. Therefore, social capital resulting in collaboration networks can be used to explain the concept of knowledge capital (Oh et al., 2006).

In order to quantify and highlight the importance of academic collaboration activities, studies exist that measure these not only using bibliometric indicators such as Rc-Index (Abbasi et al., 2010), but also social network measures (Abbasi & Altmann, 2011; Abbasi, Altmann, & Hossain, 2011; Takeda, Truex III, & Cuellar, 2010; Yan & Ding, 2009; Zhuge & Zhang, 2010). These studies have shown the applicability of social network measures (e.g., centrality measures) for co-authorship networks to indicate how centrality measures (as a proxy for scholars' collaboration activity) are useful to reflect scholars' performance based on their position and influence within their collaboration network. Here, also in another attempt to assert the importance of co-authors' role and position in their collaboration network, we evaluate a co-authorship network and propose measures for scholars' social capital. We highlight the proposed measure which better reflects scholars' social capital having higher correlation with their citation-based performance (i.e., citation count and h-index).

The motivating questions for our study are: (i) How do we measure the concept of social capital of scholars? (ii) Do scholars' social capital metrics associate with their performance? For our analysis at an exploratory level, we use co-authorship and citation data obtained from the Scopus bibliometric database looking for publication records having "*information science*" in their title or keywords or abstracts during the time period of 2001 and 2010. This enables us to shape a co-authorship network of active scholars in the field of "*information science*".

In the following sections, we review the existing literature on social capital and network theories leading to measures proposed for social capital and co-authorship and performance. In Section 3, we explain our data collection method followed by our methodology and proposed measures to quantify scholars' social capital. Then, the result of testing associations between scholars' social capital measures and their performance is shown in the following section. We conclude the paper by discussing our findings and research limitations in Section 5.

## 2. Social Capital and Network Theories

The concept of social capital has become increasingly popular in a wide range of social science disciplines (e.g., political science, economics, and organizational science). Social capital has been used as an important factor to explain actors' success in a number of areas (e.g., educational performance, career success, product innovation, inter-firm learning, and real-estate sales) by social scientists. Hanifan (1916) work on evaluating effect of community participation in enhancing school performance can be considered as the first study on social capital. But Bourdieu's (1986; 1992) and Coleman's (1987, 1988, 1990) works on education and Putnam's (1993, 1995, 2001) works on civic engagement and institutional performance are the main studies inspiring most of the current researches in social capital (Woolcock & Narayan, 2000).

Bourdieu (1986) identified several forms of capital: *Economic capital*; *cultural capital*: which could be embodied (in persons), objectified (e.g., in art), institutionalized (e.g., university degrees); *Social capital*: resources grounded in durable exchange-based networks of persons; *Symbolic capital*: manifestation of each of the other forms of capital when they are naturalized on their own terms. Bourdieu and Wacquant (1992) defined social capital in detail as "the sum of the resources, actual or virtual, that received by an individual (or a group) due to having a lasting network of more or less institutionalized relationships of mutual acquaintance and recognition" (p. 119).

Coleman (1988), a sociologist interested in the role of social capital in human capital creation and educational outcome (Narayan & Cassidy, 2001), defines social capital as a function of social structure producing advantage: "It is not a single entity but a variety of different entities, with two elements in common: they all consist of some aspect of social structures, and they facilitate certain actions of actors-whether persons or corporate actors-within the structure." (p. 598). Putnam (1993) also defined social capital as "those features of social organization, such as trust, norms and networks that can improve the efficiency of society by facilitating coordinated actions" (p. 167) or as "features of social life - networks, norms and trust - that enable participants to act together more effectively to pursue shared objectives" (Putnam, 1995) (pp. 664-665).

Coleman's (1988) definition regards social capital as one of the potential resources, which an actor can use besides other resources such as human or cultural capital (their own skills and expertise), physical capital (tools) and economic capital (money) (Gauntlett, 2011). He also highlighted the importance of social capital effecting the creation of human capital. But social capital differs fundamentally from other types of capital as it resides not in the objects themselves (e.g., people) but in their relations with other objects. For instance, human capital represents individual attributes and characteristics (e.g., attractiveness, intelligence, and skills). These assets are possessed by individuals but social capital is embedded in the relationships among individuals (Shen, 2010).

Emphasizing social capital's function in different contexts, Portes (1998) defines social capital as "the ability of actors to secure benefits by virtue of memberships in social networks or other social structures" (p. 3). Furthermore, Adler and Kwon (2002) defined social capital as "the resources available to actors as a function of their location in the structure of their social relations" (p. 18). They focus on social capital as a resource that exists essentially (permanently) in the social network binding a central actor to other actors.

In another approach, Lin (1982) in the author's social resource theory claimed power, status and wealth as determinants of *valued resources* in most societies. Accessing and using social resources can lead to better socioeconomic status and are determined by structural positions and use of ties. In addition, some researchers defined social capital considering capital (attributes) individuals posses in a network. For instance, Boxman et al. (1991) defined social capital as "the number of people who can be expected to provide support and the resources those people have at their disposal" (p. 52) and Burt (1992) defined as this concept as "friends, colleagues, and more general contacts through whom you receive opportunities to use your financial and human capital" (p. 9) and also "the advantage created by a person's location in a structure of relationships" (Burt, 2005) (p. 5). Therefore, from this point of view, social capital can be evaluated by the amount or variety of such characteristics of other actors to whom an actor has ties directly or indirectly (Lin, 1999). The core idea is that the actions of individuals (and groups) can be greatly facilitated by their direct and indirect links to other actors in their respective social networks (Adler & Kwon, 2002).

In the above definitions, the focus is on the sources (e.g., networks, norms and trust) rather than the consequences of social capital. They considered different dimensions for social capital, namely bonding and bridging (Woolcock & Narayan, 2000) taking into account valued social resources. On the bonding views of social capital, the focus here is on collective actors' internal characteristics and ties structure (Adler & Kwon, 2002). Therefore, the bonding view of social capital undergirds reciprocity and solidarity, builds trust within the group and provides substantive and emotional support (Shen, 2010). Bonding social capital is viewed as a property of a network (group of individuals), which is not the focus of our study.

## 2.1. Individual's Social Capital-Related Theories of Network

### 2.1.1. Tie Strength Theories

Granovetter (1973)'s theory of the 'strength of weak ties' argues that an individual obtains new and novel information from weak ties rather than from strong ties within the individual's group structure. Examining people looking for a job, Granovetter (1973) illustrated that there were two kinds of social relationships: weak ties and strong ties. Contrary to popular belief, he found that the most successful job seekers were not those with the strongest ties. On the contrary, because weak ties with acquaintances provide a broader set of information and opportunities, they are more helpful during people's job search than strong ties with family and friends.

The strength of a link between actors (interpersonal tie) in a network could be indicated and measured by the amount of time the link has been established, the degree of emotional intensity, the degree of intimacy, and reciprocal services (Granovetter, 1973). The interaction among the individual creates opportunity for knowledge sharing and information exchange and is considered crucial in the building trust among individuals.

On the other hand, Krackhardt (1992) showed that strong ties are important in the generation of trust. He introduced the theory of 'strength of strong ties' in contrast to Granovetter's (1973) theory. Levin and Cross (2004) found that strong ties, more so than weak ties, lead to the receipt of useful knowledge for improving performance in knowledge-intensive work areas. However, controlled for the dimension of trust, the structural benefit of weak ties emerged in their research model. It suggests that the weak ties provide access to non-redundant information. Weak ties facilitate faster project completion times, if the project is simple. It enables faster search for useful knowledge among other organizational subunits. Strong ties foster complex knowledge transfer, if knowledge is highly complex (Hansen, 1999; Reagans & Zuckerman, 2001).

**2.1.2. Structural Hole Theory**

Burt (1992) argues that the structural configuration of an individual's social network, which provides optimized "bridging" or "brokerage" position is what dictates structural advantages such as information novelty and control. The basis for this argument leverages on the fact that maximizing the number of ties (ego-network size), regardless of being weak or strong, in an individual's network does not necessarily provide benefits. Furthermore, as an individual's personal network grows over time, the extent of information coming from closely knit clusters tends to become redundant.

This is consistent with Freeman's (1979) approach to betweenness which is build around the concept of 'local dependency'. Therefore, it could be seen that Burt's (1992) notion of structural holes built further upon the assumption of betweenness centrality that advocated the idea of a brokerage position as providing information and control benefits. In fact, he shifted focus from the network structure to network position (Chung & Hossain, 2009).

Burt (1992) claimed that increasing the number of direct contacts (ego-network size) without considering the diversity reached by the contacts makes the network inefficient in many ways. Therefore, the number of non-redundant contacts is important to the extent that redundant contacts would lead to the same people and, hence, provide the same information and control benefits. He defined ego-network *effectiveness* as the number of clusters which the ego is connected to and can obtain novel information and benefits (Burt, 1992).

A structural hole (hole in the network structure) is defined as lack of tie between any pair of actors in the network. Network brokerage refers to the social structure where an actor builds connections across structural holes (Burt, 2005) linking otherwise disconnected actors. Brokerage brings novel information and opportunities, but the connections are too weak to provide emotional and substantive support. For instance, in economic networks, producers that broker more structural holes were found to make better profits from negotiating more favorable transactions with suppliers and customers (Burt, 1992). Within organizations, individuals' mobility is enhanced by having an informational network rich in structural holes (Podolny & Baron, 1997).

Thus, Burt (1992) capitalizes on his theory of structural holes by focusing on the importance of structural position (e.g., brokerage) rather than structural properties (e.g., ego's network size). Therefore, on this view of social capital as bridging, social capital can help explain the differential success of actors (e.g., individuals and firms). Therefore, bridging social capital leads to a broad worldview, diversity in opinions and resources, and information diffusion (Shen, 2010). Bridging view of social capital focuses on a property of individuals (ego-network and not whole-network).

These views highlight the social network engagement as a prerequisite for social capital. Walker et al. (1997) highlighted that "a social network structure is a vehicle for inducing cooperation through the development of social capital" (p. 110). Therefore, in brief, social capital could be regarded as the value of social networks, bonding similar people and bridging between diverse people, with norms of reciprocity (Uslaner, 2001).

## 2.2. Measuring Individuals' Social Capital

Measuring social capital is required in order to use it as a development tool. Although multi-dimensionality (i.e., different levels and units of study) and dynamicity of social capital over time (due to change of the social interaction over time) makes obtaining a single, true measure almost impossible (Woolcock & Narayan, 2000) but several researchers proposed different metrics to measure social capital.

Bourdieu's (1986) tool to quantify social capital is network size: "The volume of the social capital possessed by a given agent thus depends on the size of the network of connections he/she can effectively mobilize and on the volume of the capital (economic, cultural or symbolic) possessed in his/her own right by each of those to whom he/she is connected" (p. 249). We should consider that while greater network size is good but the quality of the individuals is crucial for social capital.

As explained earlier, it could infer that social capital is rooted in social networks and social relations and must be measured relative to them (Lin, 1999). Therefore, network science and social network analysis metrics could be used for measuring social capital. In this regard, several researchers asserted the location of actors in a network: ties strength (Granovetter, 1973; Portes, 1998), structural hole and constraints (Burt, 1992), as the key element of identifying social capital.

As social network's engagement is the principal for social capital, we also use social network analysis metrics (that support the dimensions discussed in the literature) to measure social capital of scholars in their co-authorship network. We summarize the main indicators followed by their focus as discussed in the literature in Table 1.

Table 1. Social Capital Dimensions and Relevant Proposed Metrics as Assets in Network

| Indicators | Focus | Authors |
|---|---|---|
| **Ego network size** | Diversity of contacts | (Bourdieu, 1986) & (Boxman et al., 1991) |
| **Ego Average Ties Strength** | Ties Strength | (Granovetter, 1973) & (Krackchardt, 1992) |
| **Ego Betweenness Centrality** | Structural Position | (Burt, 1992) |
| **Ego Effectiveness** | Brokerage & diversity | (Burt, 1992) |
| **Contact Status (Power)** | Embeddedness resources | (Lin, 1982) & (Burt, 2005) |

To measure social capital, we use indicators covering the focus of related studies. Diversity of contacts (ego's network size) (Bourdieu, 1986), representing the available resources for an individual, has been considered in the literature as one of important factors on information diffusion and novelty. Another important factor emphasized in the literature is ties strength (Granovetter, 1973; Krackhardt, 1992). In addition, we use ego-betweenness centrality (Freeman, 1979) and effectiveness (Burt, 1992) in order to measure the structural position and brokerage characteristics of an individual in the network, respectively.

In order to provide synthesis of two different approaches of social capital, diversity and power (as determinants of valued resources) (Lin, 1982), we define social capital as the frequency and diversity of contacts (directly connected actors) to the powerful (high performance) contacts. Therefore, having the power (value) of actors in a social network, we measure social capital for an individual. We propose a new measure, Power-Diversity Index, to take into consideration the value of direct contacts in addition to their (quantity). Furthermore, proposing another measure, Power-Tie-Diversity Index, we added tie strength factor also to the previous measure (Power-Diversity Index). These measures are explained in detail in Section 3.2. These new proposed measures (i.e., Power-Diversity, Power-Tie-Diversity) reflect the thinking that connecting to more powerful individuals will give individuals more power. Accordingly then, this reflects individuals' power and influence on transmitting and controlling information as well as the popularity of an individual based on popularity of its direct contacts.

## 3. Data, Method and Measures

### 3.1. Data

Scopus is one of the main sources presenting bibliometric data. To construct our database, we extract publications using the phrase "*information science*" in their titles or keywords or abstracts and restricting the search to publications in *English* published between *2001* and *2010*. Indeed, the publications extracted cannot be considered as representing the world production in the "*information science*" field but it illustrates a good portion of publications in this field that do not have limitation to a specific sub-field, conference, journal, institutes and country.

After extracting the publications' meta-data from Scopus and importing the information (i.e., title, publication date, author names, affiliations, publisher, number of citations, etc.), we used an application program, for extracting relationships (e.g., co-authorships) between and among researchers, and stored the data in tables in a local relational database. Four different types of information were extracted from each publication meta-data: Publications information (i.e., title, publication date, journal name, etc.); authors' names; affiliations of authors (including country, institute and department name, etc.); and keywords.

Exploring our original extracted data we found affiliation information inconsistent, where there were several fields missing for some of publications and also different written names for the country of origin and institutions. So, in our second step we undertook manual checks (using Google) to fill the missing fields using other existing fields (e.g., we used institute names to find country). Also manually we merged the universities and departments which had different names (e.g., misspellings or using abbreviations) in our original extracted. Finally, after the cleansing of the publication data, the resulting database contained 4,579 publications published in 1,392 journals and conference proceedings (Indexed by Scopus) reflecting the contributions of 10,255 authors from 99 countries.

### 3.2. Methodology

Social network analysis (SNA) is the mapping and measuring of relationships and flows between nodes of a social network. SNA provides both a visual and a mathematical analysis of human-influenced relationships. The social environment can be expressed as patterns or regularities in relationships among interacting units (Wasserman & Faust, 1994). Each social network can be represented as a graph made of nodes (e.g., individuals, organizations, information) tied by one or more specific types of relations, such as financial exchange, friends, trade, and Web links. A link between any two nodes exists, if a relationship between those nodes exists. For instance, if the nodes represent people, a link means that those two people know each other in some way.

Measures of SNA, such as centrality, have the potential to unfold existing informal network patterns and behavior that are not noticed before (Brandes & Fleischer, 2005). A method used to understand networks and their participants is to evaluate the location of actors in the network. Measuring the network location is about determining the centrality of an actor. These measures help determine the importance of a node in the network. To quantify the importance of an actor in a social network, various centrality measures have been proposed over the years (Scott, 1991; Wigand, 1988).

Using each publication and its authors, we construct the co-authorship network of scholars. Nodes of the network represent scholars and a link between two nodes represents a publication co-authorship relationship between or among those scholars. We used UCINET (Borgatti et al. 2002), a social network analysis tool for visualizing the network and statistical functions, for calculating the measures descried below.

### 3.3. Measures
### 3.3.1. Measuring Scholars Performance

To assess the performance of scholars, many studies suggest quantifying scholars' publication activities as a useful measure for the performance of scholars. But there are also many researchers pointing to the limits and bias of such quantification focusing on publication, mainly on the most visible articles from international databases. Further research shows the number of citations a publication receives qualifies the quantity of publications (Lehmann, Jackson, & Lautrup, 2006). Cronin (1996; 2002) also has been emphasized in the accumulation of citation counts, a form of symbolic capital, as an important aspect of academic life. Progressively, new citation-based metrics are being proposed, following Hirsch's (2005) h-index as the core metric for measuring the combination of quantity and quality of researchers and academic communities. Although there is considerable debate on the reliability of the h-index (e.g., Haque & Ginsparg, 2009), the h-index is still widely used world-wide among academicians. While the reliability of the measure is not the subject of this paper per se, it does provide at least an empirical and very widely used metric so as to gauge a researcher's prolificacy. Thus, we will consider the h-index as a citation-based surrogate measure and as a proxy for the performance of scholars.

### 3.3.2. Measuring Scholars' Social Capital

In Table 1 and the following paragraphs, we demonstrate the proper indicators recommended in the literature to measure social capital. We explain each indicator definition and their respective equations more precisely in this section. To answer precisely our first research question: "how do we measure the social capital of scholars?" we propose following metrics to measure individuals' social capital. Although some of them have been used previously but we propose two new measures (*Power-Diversity Index* and *Power-Tie-Diversity Index*) which combines two and three different properties of individuals in their social network in order to quantify their social capital.

- *Individual Network Size (Degree Centrality)*

In order to measure diversity of contacts representing the available resources for an individual, as one of important factors of information diffusion and novelty, we will use individual degree centrality which is the number of direct contacts it has. In a co-authorship network, network size of an author is the number of her co-authors.

- *Individual Ties Strength (Average Tie Strength & Weighted Degree Centrality)*

To evaluate an individual's ties strength, we use the sum of ties strength and also average ties strength as proxy for social capital in order to represent the average strength of each tie of the actor.

Sum of ties strength of an author is the total number of collaborations she has (including redundant collaborations with any co-author). Average ties strength is simply the average of the weights of her collaborations. That means dividing the sum of ties strength (i.e., the number of collaborations) by the network size of the author (i.e., the number of different co-authors).

- *Individual Effectiveness*

In order to optimize an individual's network by capitalizing on structural holes, Burt (1992) claims that increasing the number of direct contacts *(network size)* without considering the diversity reached by the contacts makes the network inefficient in many ways. Therefore, the number of non redundant contacts is important to the extent that redundant contacts would lead to the same people and hence provide the same information benefits. The term *effectiveness* is used to denote the average number of people reached per primary contact and to denote effectiveness in networks. Burt (1992) uses 'effective size' as a term to denote the same.

In conclusion, effectiveness of an individual is defined as the number of non-redundant (not connected) contacts. Precisely, this is the number of contacts that an individual has, minus the average number of ties that each contact has to other contacts of individuals. In a co-authorship network, effectiveness of an author is the number of her co-authors which are not co-author with each other.

- *Ego-Betweenness Centrality*

Considering bridging dimension, we use actors' ego-betweenness centrality to measure social capital. Betweenness centrality is an indicator of an individual's potential control of communication within the network and highlights bridging (brokerage) behavior of an actor (Freeman, 1979). Ego-betweenness centrality is defined as the sum of an individual's proportion of times this individual lies on the shortest path between each part of alters (direct contacts to ego) (Hanneman & Riddle, 2005). For alters connected to each other, the contribution to the ego-betweenness of that pair is 0, and for contacts connected to each other only through ego (individual), the contribution is 1, for alters connected through ego and one or more other alters, the contribution is 1/k, where k is the number of nodes which connects that pair of alters.

- *Individual Power-Diversity Index*

In order to synthesize the two different approaches of social capital, diversity and power, we define the individual Power-Diversity Index to measure social capital based on both the frequency of connections and also considering the power of contacts (directly connected individuals). Having the power (value) of individuals in a social network, we could simply calculate an individual's sum or average of the power of direct contacts to synthesize quantity (frequency of contacts) and quality (their value) of embedded resources (contacts) of an individual as a proxy for his/her social capital. But in order to have a more advanced and accurate metric (rather than merely the sum or average), we will use the h-index (Hirsch, 2005) base formula to quantify the quality of contacts of an individual by counting top h powerful (valued) contacts whose power value is at least h.

In a co-authorship network, we consider the h-index of authors as their power (value) indicator. Therefore, the *Power-Diversity Index* of an individual is the largest number such that her top *h* co-authors have each at least an h-index of *h*. For instance, looking at Table 2 the author has 7 co-authors who have h-indices of 6, 5, 5, 3, 3, 2 and 1, her *Power-Diversity Index* is 3 as three of her co-authors have an h-index of higher than 3 and one cannot find 4 co-authors who have an h-index of higher than 4.

- *Individual Power-Tie-Diversity Index*

In another effort, we take into consideration also individuals' tie strengths as another important property of individuals' social capital discussed in the literature. This measure can be applied in weighted networks. It is similar to individual *Power-Diversity Index* but taking the weight (strength) of ties into account. To define this new measure for an individual (in a weighted network), first we define co-authors' *power-strength* which is the h-index of each co-author multiplied by the strength of the tie between that co-author and the author. So, individual *Power-Tie-Diversity Index* is the largest number such that her top *h* co-authors have each at least the *power-strength* of *h*.

**Table 2. An Individual's co-authors and their h-index and frequency of collaborations**

|   | Co-authors | h-index | Freq. of Collaborations | power-strength |
|---|---|---|---|---|
| 1 | CA1 | 6 | 4 | 24 |
| 2 | CA2 | 5 | 3 | 15 |
| 3 | CA3 | 5 | 2 | 10 |
| 4 | CA4 | 3 | 3 | 9 |
| 5 | CA5 | 3 | 1 | 3 |
| 6 | **CA6** | 1 | 2 | 2 |
| 7 | **CA7** | 1 | 2 | 2 |

In a co-authorship weighted network to calculate the individual *Power-Tie-Diversity Index* first we need to calculate the *power-strength (co-ps)* of each of her co-authors as her h-index times the number of collaborations (tie strength) they have had. Then, *Power-Tie-Diversity* of an individual is the largest number such that her top *h* co-authors have each at least *co-ps* of *h*. For instance, Table 2 shows the co-authors' power-strength of an author which are 24, 15, 10, 9, 3, 2, and 2 in descending order. Thus, the author's *Power-Tie-Diversity Index* is 4 as for 4 of her co-authors' power-strengths (*co-ps*) are higher than 4.

## 4. Analysis and Results

### 4.1. Dataset statistics

Table 3 shows the top 20 active journals which have more publications in "*information science*" (having "information science" in the title or keywords or abstracts). Table 4 indicates the top 20 journals based on the citations count each journal receives.

Table 3. Top 20 active journals in the field of "information science"

| | Journal Name | Pub. | Cit. | | Journal Name | Pub. | Cit. |
|---|---|---|---|---|---|---|---|
| 1 | Information Sciences | 229 | 2818 | 11 | Aslib Proceedings: New Info. Perspect. | 33 | 43 |
| 2 | JASIST | 223 | 3297 | 12 | Physical Review A | 32 | 638 |
| 3 | Journal of Information Science | 98 | 1195 | 13 | Library Trends | 32 | 64 |
| 4 | J. of Automation and IS | 66 | 8 | 14 | Bioinformatics | 29 | 1193 |
| 5 | Journal of Documentation | 60 | 387 | 15 | Information Research | 29 | 168 |
| 6 | J. of Chemical Info. and Modeling | 54 | 627 | 16 | Library Review | 29 | 27 |
| 7 | Education for Information | 51 | 125 | 17 | Journal of Theoretical Biology | 28 | 617 |
| 8 | Library and IS Research | 50 | 289 | 18 | New Library World | 27 | 48 |
| 9 | International J. of Medical Info. | 40 | 248 | 19 | International Info. and Library Review | 27 | 39 |
| 10 | WSEAS Transactions on IS & App. | 37 | 22 | 20 | Physical Review Letters | 26 | 1595 |

Table 4. Top 20 cited journals in the field of "information science"

| | Journal Name | Cit. | Pub. | | Journal Name | Cit. | Pub. |
|---|---|---|---|---|---|---|---|
| 1 | Nature | 3577 | 18 | 11 | PNAS | 734 | 24 |
| 2 | JASIST | 3297 | 223 | 12 | Nucleic Acids Research | 661 | 17 |
| 3 | Information Sciences | 2818 | 229 | 13 | Physical Review A | 638 | 32 |
| 4 | Science | 1934 | 22 | 14 | J. of Chemical Info. and Modeling | 627 | 54 |
| 5 | Physical Review Letters | 1595 | 26 | 15 | Journal of Theoretical Biology | 617 | 28 |
| 6 | Journal of Information Science | 1195 | 98 | 16 | Nature Biotechnology | 538 | 4 |
| 7 | Bioinformatics | 1193 | 29 | 17 | Genome Research | 527 | 2 |
| 8 | Information Systems Research | 850 | 7 | 18 | SIGMOD Record | 476 | 12 |
| 9 | MIS Quarterly | 784 | 12 | 19 | Nature Physics | 441 | 13 |

| 10 | Reviews of Modern Physics | 749 | 2 | 20 | JAMIA | 393 | 22 |

## 4.2. Scholars Performance and Social Capital Measures

Based on the available publication meta-data of scholars, we retrieve every pair of authors who are listed as authors of a publication. We merge redundant co-authorships by increasing more weight (tie strength) to their link (tie) for each relation. Therefore, we form the co-authorship network of scholars and a weighted network. This relational data (i.e., who is connected to whom with which frequency) is the basis for social network analysis. We imported these data to UCINET (Borgatti, Everett, & Freeman, 2002) to calculate the social network measures. In addition we calculated the citation-based performance measures (i.e., h-index) for all scholars. To illustrate, the results for the top 10 productive scholars are shown in Table 5.

Table 5. Top 10 high performance scholars and their social capital measures

|   | Name | h-index [1] | Cit. Cnt. | Network Size (Degree) | Weighted Degree | Effectiveness | Avg. Ties Strength | Ego-Betweeness | Pow-Div | Pow-Tie-Div |
|---|---|---|---|---|---|---|---|---|---|---|
| 1 | M. Thelwall | 9 | 460 | 17 | 26 | 16.45 | 1.53 | 245 | 4 | 2 |
| 2 | H.J. Kimble | 8 | 1125 | 28 | 40 | 23.68 | 1.43 | 557.3 | 4 | 3 |
| 3 | Y. Wang | 8 | 328 | 30 | 35 | 19.56 | 1.17 | 376 | 3 | 2 |
| 4 | E.R. Dougherty | 7 | 606 | 16 | 21 | 13.88 | 1.31 | 186 | 4 | 2 |
| 5 | B. Cronin | 6 | 164 | 4 | 6 | 4.33 | 1.50 | 12 | 2 | 1 |
| 6 | C. Oppenheim | 6 | 153 | 20 | 26 | 19.19 | 1.30 | 352 | 4 | 2 |
| 7 | L.I. Meho | 5 | 282 | 6 | 8 | 5.67 | 1.33 | 24 | 1 | 1 |
| 8 | H.D. White | 5 | 169 | 2 | 2 | 1.00 | 1.00 | 0 | 1 | 1 |
| 9 | J.C. Principe | 5 | 120 | 13 | 14 | 10.79 | 1.08 | 114 | 1 | 1 |
| 10 | Y.B. Jun | 5 | 89 | 6 | 6 | 5.33 | 1.00 | 26 | 1 | 1 |

To answer our second research question: "Do scholars' social capital metrics associate with their performance?" we calculated all measures shown in Table 5 for all scholars. Then, we applied the Spearman correlation rank test between the social capital measures and scholars' performance (i.e., citations count and h-index). As Table 6 shows the results of the correlation test, there are high significant correlation coefficients between social capital measures and scholar's citation-based performance. Results suggest that individuals' *Power-Diversity* Index has the highest coefficient with their performance either considering citation count or h-index. This highlights the importance of the power and role of co-authors to generate social capital for an author in her co-authorship network which may also lead to enhance her performance.

Table 6. Spearman correlation rank test between scholars' centrality measures and their performance

| Scholars' Social Capital Measures (N=10,254) | Scholars' Performance Measure |
|---|---|

|  | Citation Count | h-index |
|---|---|---|
| **Individual Network Size (Degree Cent.)** | .219 * | .159 * |
| **Weighted Degree Centrality (Sum of Ties Strength)** | .226 * | .177 * |
| **Average Ties Strength** | .135 * | .268 * |
| **Individual Effectiveness** | .192 * | .292 * |
| **Ego Betweenness Centrality** | .172 * | .309 * |
| **Individual Power-Diversity Index** | **.656 *** | **.853 *** |
| **Individual Power-Tie-Diversity Index** | .103 * | .206 * |

*. Correlation is significant at the .01 level (2-tailed).

It is noteworthy to point out that the *Power-Diversity* Index coefficient value is three times higher than the second highest measure (i.e., ego-betweenness for the h-index and weighted degree centrality for citation count). Interestingly the coefficient for *Power-Tie-Diversity* Index is much lower than the *Power-Diversity* Index. This indicates that repeated collaborations with the same co-authors (even if they are prominent) does not create good social capital for them rather than having collaborations with more powerful (prominent) co-authors.

The Ego-betweenness centrality coefficient is higher than ties strength and diversity measures. This shows bridging characteristics of scholars in their co-authorship network seems more important than the diversity of their co-authors and their ties' strength in regard to their performance.

## 5. Conclusion and Discussion

In this study, we highlighted the importance of the co-authorship network as a tool for evaluating scholars' performance which is necessary in academia. We use social capital theory to explain how scholar's co-authorship network affects each individual scholar's performance. Although there are several definitions for social capital, most definitions' emphases are on the social relations that have productive benefits. Social capital is rooted in social networks and social relations and must be measured relative to them (Lin, 1999). Reviewing the literature we highlight scholars' structural position (e.g., network size, degree, betweenness) in their co-authorship network and also scholars' contacts characteristics (power and performance) as proper indicators of their respective social capital.

Reviewing the literature on social scientists' and particularly network scientists' studies on social capital, we highlight different approaches and dimensions for social capital and focus on the approaches which evaluate individuals' property in the network. Although multi-dimensionality and dynamicity of social capital makes having a single, true measure almost impossible (Woolcock & Narayan, 2000) but as measuring social capital is required in order to use it as a development tool, several researchers proposed different metrics.

Several measures (i.e., individual network size, ties strength, ego-betweenness centrality, *Power-Diversity* Index and *Power-Tie-Diversity* Index) using network analysis metrics. This assists us in quantifying the social capital resulting from the co-authorship through the social network, which is considered important for research management, academic institute as well as government policy makers over recent years. While several measures have been used by other researchers earlier, the last two measures (i.e., *Power-Diversity Index* and *Power-Tie-Diversity Index*) are new which combine two and three properties of authors in their co-authorship network, respectively, to quantify the extent of social capital they gain.

We test the correlation of the proposed measures of authors' with their research performance and all show a positive significant association. The results highlighted the importance of scholars' social capitals characteristics on their performance. Significant association between scholars' *Power-Diversity Index* and performance follows that connecting to more powerful contacts will lead to have better performance which is due to contacts' relative power and influence on transmitting and controlling information as well as the popularity of an individual based on popularity of its direct contacts. *Power-Diversity Index* indicates the individuals diversely connected to prominent contacts. These kinds of actors have special strategically positions that can control the flow of information in the network.

Our research conceptualizes social capital in terms of network structures, such as articulated by the strength of weak ties theory (Granovetter, 1973, 1982) and provides valuable insight into co-authors' activities. The strength of weak ties theory suggests that the social network of any network member is the co-author's primary resource. Moreover, this network can be viewed as being comprised of participants who vary by the relative strength of their relationship with one another. Strongly tied-together members in a network tend to be more similar to each other than different, more likely to be available for each other, share more common interests, and interact more frequently. Conversely, weakly tied members in a social network tend to communicate less frequently, are more different than similar, and provide both newer information into the network and more access to other social networks (Sawyer, Crowston and Wigand, 1999). When applied to the co-author network, this suggests that co-authors with large social networks populated with more weak ties will have more social capital. The more resource-rich co-authors will get influential linkages and connections (via acquaintances) and be able to point to more influential co-authors who might be able to provide value-adding services.

In brief, our findings show that *Power-Diversity* is a useful surrogate of the importance of a scholar in her co-authorship network by considering the diversity of contacts (number of co-authors) and also their value and power (performance). *Power-Diversity,* which can be considered as a new hybrid centrality measure, identifies individuals having direct connections to diverse powerful individuals. So, this measure is an indicator of the power and influence of an individual's ability to control communication and information.

Using publication data and extracting co-authorship relations, we have presented an overview of collaboration efforts and collaborative networks in the "*information science*" research area. The collaboration networks of scientists in "*information science*" have been analyzed by using author affiliations from publications having '*information science*' in their 'title' or 'keywords' or 'abstract' since 2001 to 2010 as extracted from Scopus. The publication dataset we have extracted does not support to represent the complete world production of research on "*information science*" (due to the possibility of significant biases: ignoring the relevant publications which are not using the exact phrase

'*information science*'; publications in other languages). Hence, the database does not pretend to represent the complete field.

Applying this new measure for other social networks to test its association with individuals' performance could be a useful extension of standard centrality measures and a suitable proxy for the performance of the individuals in a network. In order to accomplish this, validation of this new measure is needed by testing it in other social networks.

**Footnotes**:

[1] The h-indices of scholars are not their full h-index. It is calculated based on their ten years of publications extracted from Scopus for our query between 2001 and 2010. This is the reason of difference with the h-indices of scholars reported in (Cronin & Meho, 2006).